%% file: pattern_improvement.tex
\documentclass[%
 article
nofootinbib,
 amsmath,amssymb,
 aps,
 pra,
 onecolumn,
 notitlepage
]{revtex4-1}

\usepackage[centering,hmargin=2cm,vmargin=2.5cm]{geometry}

\usepackage{graphicx}	
\usepackage{dcolumn}	
\usepackage{bm}			

\usepackage{braket}
\usepackage{listings}
\usepackage{color}
\usepackage{enumitem}   	
\setlist{nolistsep}

\usepackage{slashbox}			

\usepackage{algorithm}
\usepackage{algorithmicx}
\usepackage{algpseudocode}
\algnewcommand\algorithmicswitch{\textbf{switch}}
\algnewcommand\algorithmiccase{\textbf{case}}
\algnewcommand\algorithmicassert{\texttt{assert}}
\algnewcommand\Assert[1]{\State \algorithmicassert(#1)}%
\algdef{SE}[SWITCH]{Switch}{EndSwitch}[1]{\algorithmicswitch\ #1\ \algorithmicdo}{\algorithmicend\ \algorithmicswitch}%
\algdef{SE}[CASE]{Case}{EndCase}[1]{\algorithmiccase\ #1}{\algorithmicend\ \algorithmiccase}%
\algtext*{EndSwitch}%
\algtext*{EndCase}%

\definecolor{codegreen}{rgb}{0,0.6,0}
\definecolor{codegray}{rgb}{0.5,0.5,0.5}
\definecolor{codepurple}{rgb}{0.58,0,0.82}
\definecolor{backcolour}{rgb}{0.95,0.95,0.92}

\def\note#1{}
\def\extra#1{}

\def\L{\mathcal{L}}		
\def\G{\mathcal{G}}		
\def\A{\mathcal{A}}		
\def\N{\mathcal{N}}		
\def\R{\mathcal{R}}		
\def\sc{\mathcal{S}}	
\def\I{\mathcal{I}}		
\def\C{\mathcal{C}}		

\lstdefinestyle{mystyle}{
    backgroundcolor=\color{backcolour},   
    commentstyle=\color{codegreen},
    keywordstyle=\color{magenta},
    numberstyle=\tiny\color{codegray},
    stringstyle=\color{codepurple},
    basicstyle=\footnotesize,
    breakatwhitespace=false,         
    breaklines=true,                 
    captionpos=b,                    
    keepspaces=true,                 
    numbers=left,                    
    numbersep=5pt,                  
    showspaces=false,                
    showstringspaces=false,
    showtabs=false,                  
    tabsize=2
}
 
\begin{document}

\title{Scheduler of quantum circuits based on dynamical pattern improvement\\
and its application to hardware design}

\author{Gian Giacomo Guerreschi}
 \email{gian.giacomo.guerreschi@intel.com}
\affiliation{Intel Labs}

\date{\today}


\begin{abstract}
As quantum hardware increases in complexity, successful algorithmic execution relies more heavily on awareness of existing device constraints.
In this work we focus on the problem of routing quantum information across the machine to overcome the limited connectivity of quantum hardware. Previous approaches address the problem for each two-qubit gate separately and then impose the compatibility of the different routes. Here we shift the focus onto the set of all routing operations that are possible at any given time and favor those that most benefit the global pattern of two-qubit gates. We benchmark our optimization technique by scheduling variational algorithms for transmon chips. Finally we apply our scheduler to the design problem of quantifying the impact of manufacturing decisions. Specifically, we address the number of distinct qubit frequencies in superconducting architectures and how they affect the algorithmic performance of the quantum Fourier transform.
\end{abstract}

\pacs{Valid PACS appear here}

\maketitle



\input{section_1_introduction} 

\input{section_2_input-output} 

\input{section_3_snapshot} 

\input{section_4_protocol} 

\input{section_5_dynamical-pattern-improvement} 

\input{section_6_hardware} 

\input{section_7_numerical-evaluation} 


\section{Conclusion and outlook}
\label{sec:conclusion}

In this work we propose a novel optimization technique to reduce the number of routing operations required to schedule a quantum algorithm for execution on realistic quantum devices. Our scheduler considers both machine-independent constraints, like the logical dependencies of the circuit's gates and the exclusive activation of qubits, and machine-dependent ones. The latter constraints are related to the limited connectivity of the physical qubits or are imposed by the control electronics. The scheduler overcomes the limited connectivity by inserting routing operations in the form of SWAP gates compatible with the electronic constraints.

The choice of which SWAP leads to a more efficient routing is based on the novel policy called dynamical-pattern-improvement. The intuitive idea is that a qubit-exchange operation may affects the distance of two 2-qubit gates among those to be scheduled next. Multiple exchanges can take place at the same time, and indeed they should in any schedule exploiting gate parallelism, with the effect of altering the distance between multiple pairs of qubits in a dynamical way. Our optimization method is aware of the global qubit placement at every moment and can select the SWAP gates that improve the pattern maximally. It is important to notice that the scheduler uses the global knowledge of the current 2-qubit pattern, without any additional look-ahead strategy. The possibility of explicit look-ahead strategies is an extension of the dynamical-pattern-improvement policy.

The efficacy of a scheduler can be quantified in terms of the final circuit depth or number of required routing operations. We benchmark our scheduler with two classes of circuits: variational quantum algorithms (QAOA) and the quantum Fourier transform (QFT). In the first case we demonstrated the utility of a simple heuristic solver of the maximum common edge subgraph to improve the initial qubit placement, together with the efficacy of the dynamical-pattern-improvement. We also quantified the importance of considering the freedom of gate reordering when gates pair-wise commute. In the second case, we explored the application of schedulers as tools to guide design decisions for device architecture. We do so by quantifying three related, but distinct, designs in terms of their algorithmic performance for QFT. We expect that this use-case will gain increasing relevance in the near future when additional manufacturing effort must be justified by a reduction of circuit depth or execution time.

Finally, the results of scheduling QAOA algorithms suggest that our stochastic optimization will profit from the division of the LDG in subgraphs, for example based on the gate priority, together with a process to progressively augment the best schedule for the smaller subgraph to that of the complete LDG. Investigations in this direction are left to future works.


\begin{acknowledgments}
The author would like to thank Adam Holmes and Nicolas P. D. Sawaya for constructive comments on an earlier version of the manuscript.
\end{acknowledgments}


\bibliographystyle{unsrt}
\bibliography{references}


\appendix
\renewcommand\thefigure{\thesection.\arabic{figure}}
\renewcommand\thetable{\thesection.\arabic{table}}
\setcounter{figure}{0}
\setcounter{table}{0}

\bigskip\noindent\makebox[\linewidth]{\resizebox{0.3333\linewidth}{1pt}{$\bullet$}}\bigskip


\section{Heuristic for maximum subgraph isomorphism}
\label{app:sec:subgraph-heuristic}

We coded a very simple heuristics to approximately solve the maximum common edge subgraph problem. Here we consider graphs $G_i(V_i,E_i)$ for $i=1,2$, each defined by a set of vertices $V_i=\{0,1,2,\dots,|V_i|-1\}$ and undirected edges $E_i=\{e_{j,k}=(v_j,v_k) \text{ s.t. } v_j,v_k\in V_i\}$. Given two graphs with $|V_1| \leq |V_2|$, one looks for the injective map $\pi:V_1\rightarrow V_2$ such that the larger possible number of edges overlap, meaning that $e_{\pi(j),\pi(k)}\in E_2$ for $e_{j,k} \in E_1$.

The heuristics starts with the choice of two vertices that we call root: $r_1 \in V_1$ and $r_2 \in V_2$. We fix $\pi(r_1)=r_2$. Then we consider the $M$ vertices at shortest distance from each of the roots. We exhaustively try all the permutations of the two sets of $M$ vertices that, added to the roots, form a $(M+1)$ subgraph homomorphism. We select the homomorphism that leads to the highest number of overlapping edges. We proceed iteratively fixing map $\pi$ on the previous set of vertices and adding the next $M$ vetices with smallest distance from the roots. In practice, we use a breadth-first-search to determine the set of $M$ vertices added in each iteration and resolve any ambiguity by selecting the vertices with lowest index. In section~\ref{sec:numerical-evaluation} we used $M=7$, $r_1=0$, $r_2=5$. Observe that with cost proportional to $|V_1|\,|V_2|$ one might explore all the possible choices of roots.


\section{Reduction of the routing overhead with native ZZ rotations for QAOA circuits}
\label{app:reduction-qaoa}

In Section~\ref{sec:numerical-evaluation} we reported a reduction up to 56\% of the number of SWAP gates added to the schedule of QAOA circuits to satisfy the connectivity constraint. Here we break-down the reduction for given number of qubits $Q_L$ and number $p$ of QAOA layers. All reduction values are provided in Table\ref{app:tab:reduction} according to the formula:
\begin{equation}
\label{app:eq:reduction}
  \text{reduction} = (1-s_\text{no-dec}/s_\text{dec}) \; ,
\end{equation}
where $s_\text{dec}$ is the number of SWAPs when ZZ rotations are decomposed for the transmon architecture and $s_\text{no-dec}$ is the number of SWAPs when ZZ rotations are native to the modified architecture, all the rest staying unchanged. The dataset are the same as plotted in Fig.~\ref{fig:qaoa-scaling} and Fig.~\ref{fig:qaoa-scaling_rzz-is-cp} of the main text.

\begin{table}[h]
\centering
\begin{tabular}{ |c||c|c|c|c|c|c| } 
 \hline
 \backslashbox{$\,p\,$}{qubits}
  & 6 & 8 & 10 & 12 & 14 & 16 \\ 
 \hline \hline
 1 & -7.3\% &  3.6\% & 22.7\% & 29.1\% & 30.8\% & 29.5\% \\
 2 &  7.6\% & 27.3\% & 39.3\% & 44.9\% & 46.0\% & 47.2\% \\
 3 & 12.1\% & 35.3\% & 44.5\% & 50.6\% & 50.8\% & 50.2\% \\
 4 & 24.2\% & 44.0\% & 50.7\% & 55.4\% & 54.3\% & 53.3\% \\
 5 & 27.7\% & 46.1\% & 52.4\% & 56.4\% & 54.8\% & 54.5\% \\
 \hline
\end{tabular}
\caption{Table quantifying the reduction in routing operations (i.e. SWAP gates) by avoiding the decomposition of ZZ rotations. The reduction is expressed as percentage from the expression in \eqref{app:eq:reduction} and refers to the QAOA circuits of section~\ref{sec:numerical-evaluation} of the main text.}
\label{app:tab:reduction}
\end{table}


\section{Logical dependency graph of QFT circuits}
\label{app:ldg-qft}

\begin{figure}[h]
\centering
\includegraphics[scale=0.31]{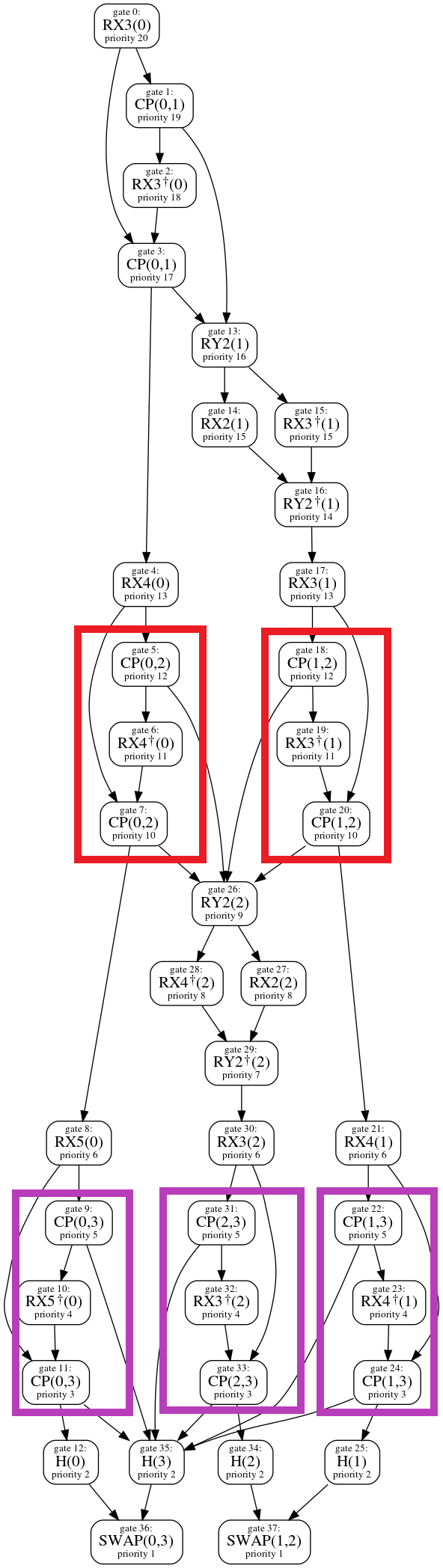}
\caption{Logical dependency graph for the QFT circuit of 4 qubits. Gates belonging to the equivalent of the ``red boxes'' of Fig.~\ref{fig:qft-circuit} (bottom panel) are surrounded by red boxes also in the LDG. Purple boxes indicate the same situation in a later position in the circuit.}
\label{app:fig:qft-ldg}
\end{figure}

To support the observations related to optimization of QFT schedules in presence of control constraints, we provide the Logical Dependency Graph corresponding to the input of our scheduler for 4 qubits. In section~\ref{sec:numerical-evaluation} of the main text we noticed that multiple two qubits gates have the same priority but the QFT scheduling profit from a specific order of execution. Fig.~\ref{app:fig:qft-ldg} makes it evident by enclosing in same-color boxes those group of gates that causes ambiguity in the scheduling. When the QFT involves more logical qubits, the number of such situations increase linearly in $Q_L$ and the occurrences becomes more complicated, involving a number of gate groups proportional to $Q_L$. With a random approach, finding the right order of execution is increasingly hard. Policy SR1 lowest-index-first removes this problem.

In addition, each box includes two 2-qubit gates: the first has higher priority than the second. This means that if we enforce the execution of gates with the highest priority, we will execute all ``first 2-qubit gate'' in each box (in random order) and then the ``second 2-qubit gate'' in each box (also in random order). This would clearly introduce a high routing overhead as compared to execute both 2-qubit gates in the same box before separating the qubit pair. Policy SR3 always-despite-priority removes this problem.

\end{document}

%% file: section_1_introduction.tex

\section{Introduction}
\label{sec:introduction}

The race towards a functioning quantum computer is pushing technological boundaries in two directions: higher-quality qubits to provide longer coherence time in absence of active error correction, and a larger number of connected qubits to increase the size and connectivity of the quantum register. To take full advantage of quantum resources in both the short and long term, these efforts must be complemented with drastic improvements in software tools.
In the short term, it is crucial to minimize algorithmic execution time in order to execute algorithms before devices decohere. In the long term, one needs efficient and effective heuristics to find near-optimal solutions to qubit routing and control, as circuits grow in size and complexity.



The technology behind quantum computers is still in its early stage and it is possible, even probable, that the first useful quantum computations will require specialized compiling tools. However, one important lesson from the computer science community is that a modular approach can be very effective in reducing the engineering effort to develop and maintain the compiler without sacrificing optimization opportunities. In the quantum computing community, the modular approach has been proposed in several works, either implicitly or explicitly \cite{Javadiabhari2014,Haener2018,Steiger2019}, and provides a framework in which different optimization techniques can be tested without the engineering overhead of a complete compiler.

In this work we focus on the problem of routing quantum information through the physical qubit register so that all two-qubit gates described in the unscheduled quantum circuit can be executed on physically connected pairs of qubits. The situation we consider is as follows: we assume that all input gates have already been decomposed into one- and two-qubit gates and define a correct schedule as one that satisfies the four conditions below. While explicitly providing the timing and parallelism of all quantum operations, a correct schedule must consider:
\begin{enumerate}
	\item the logical dependencies of the algorithm;
	\item the exclusive activation principle stating that any qubit can be involved in at most one gate at a time;
	\item the connectivity of the target architecture with two-qubit operations only between physically connected qubits;
	\item the appropriate constraints on gate parallelism due to the limitation of the control electronics.
\end{enumerate}

Here we propose a routing scheme that, given the current state of the circuit's execution, evaluates all local routing operations and prioritizes those that lead to the largest improvement in the (overall) qubit placement. This differs from previous approaches that focus on identifying the best routes to enable a single two-qubit gate and then selecting one of them based on the compatibility with the routes of other qubit pairs \cite{Dousti2012,Zulehner2018,Nishio2019,Murali2019asplos,Lao2019a}. Despite not incorporating a look-ahead or look-back strategy and therefore being temporally local, our approach naturally tends towards a spatially global optimization. We call this approach \textit{dynamical pattern improvement}.

There are several policies that can be applied to select the most beneficial routing operations. We describe a few alternatives and compare them with respect to the compilation of QAOA (a variational algorithm for finding approximate solutions of combinatorial problems) and QFT (the Quantum Fourier Transform used as subroutine in multiple quantum algorithms among which Shor's algorithm and quantum phase estimation) on bi-dimensional and linear architectures respectively. We quantify the reduction in both circuit depth and number of routing operations due to the initial qubit placement: we propose a heuristic approach based on the approximate solution of the maximum common edge subgraph problem and compared it to a randomized baseline. In addition, we compute the scaling cost of QAOA circuits with respect to the qubit number and circuit depth and observe that large optimization opportunities are opened when the commutativity of gates is exploited, leading up to 56\% reduction in the number of routing operations.

By including a full characterization of the machine it is possible to use the scheduler to guide design decisions for the architecture. In all our numerical studies, in addition to the limited connectivity we also consider the constraints imposed by the control electronics. To this end, we need to specify the hardware in detail and chose to base our study on the transmon architecture developed at QuTech in Delft together with Intel \cite{Versluis2017}.
We explore this use-case by comparing schedules of QFT circuits for linear arrays of transmon qubits that differ only in the number of distinct qubit frequencies. Our results suggest that the proposed architecture maintains a high level of parallelism and that going beyond 3 frequencies may not justify the additional manufacturing complexity.

The paper is structured as follows: Section \ref{sec:input-output} specifies the input and output of the scheduler and section \ref{sec:snapshot} introduces quantities used in the internal representation of quantum circuits. Sections \ref{sec:protocol} and \ref{sec:dynamical-pattern-improvement} describe the overall scheduling protocol and the novel optimization technique based on dynamical pattern improvement respectively. In section \ref{sec:hardware} we characterize the transmon architecture of reference \cite{Versluis2017} that is used for the numerical studies of section \ref{sec:numerical-evaluation}. Finally we draw conclusions.


%% file: section_2_input-output.tex

\section{Input and output of the scheduler}
\label{sec:input-output}


For modular compiling tools, it is important to precisely define the expected input and output of the various compiler passes \cite{DragonBook}. In our work, we assume that the quantum algorithm has already been analyzed and decomposed into a sequence of the one- and two-qubit gates that are available in the target architecture. At this point the circuit may be represented in many ways, one being simply the list of sequential gates. However, the list format does not consider the presence of commuting gates whose execution order can be safely exchanged. This is a peculiar feature of quantum circuits: the fact that certain operations acting on common qubits can be executed in arbitrary order without altering the result. To express these situations unambiguously, we prefer a graph-like structure and refer to it as the Logical Dependency Graph (LDG) \cite{Metodi2006,Guerreschi2018}.

We consider our schedule as part of the machine-dependent back-end in the compilation tool chain and this is reflected in the form of its output. As mentioned in the introduction, a correct schedule (not necessarily an optimized one, just correct) must satisfy at least four kinds of constraints. The first two (logical dependencies among the quantum operations and exclusive activation of the qubits) are arguably machine-independent, but the last two reflect the restricted connectivity of physical machines or limitations of the electronics and therefore depend on the target architecture. We provide the output in term of cQASM instructions with explicit time and parallelism. The instructions refer to operations on physical qubits.

\subsection*{INPUT: Logical Dependency Graph (LDG)}

The quantum circuit is provided as a sequential list of $G$ gates, namely $g_0, g_1, \dots g_{G-1}$. They are organized into the logical dependency graph, denoted $\L$, of which they constitute the vertices. With a natural extension of the notation, we use the same letter $g$ to indicate either the gate or the corresponding vertex of LDG. The same holds for the symbol $\G$ indicating both the set of gates and LDG vertices.
The vertices of $\L$ are connected via directed edges representing logical dependenies of the quantum algorithm, specifically the source gate must be performed before the target gate. For consistency of the algorithm, the LDG is a directed, acyclic graph.

\begin{figure}[t]
\centering
\includegraphics[scale=0.66]{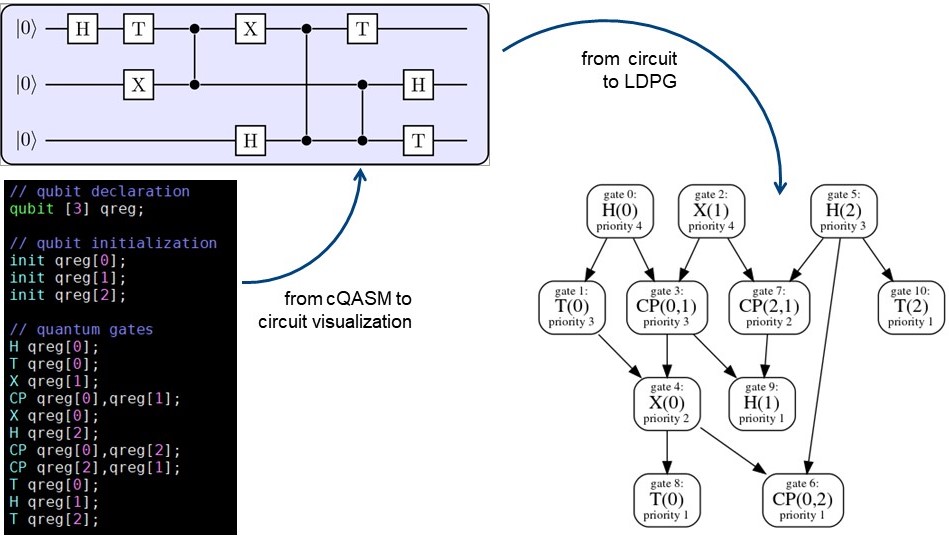}
\caption{The input of our scheduler is the logical dependency graph (LDG) of a certain quantum algorithm. Earlier in the compilation chain, the algorithm may have been provided in terms of quantum assembly instructions (leftmost panel, using the syntax known as cQASM 1.0 \cite{Khammassi2018a}), visualized as a quantum circuit (central panel), and finally translated into the LDG (rightmost panel).
In the central panel, CP gates are visualized as vertical segment joining the involved qubits. The quantum circuit includes CP and T gates that are commuting gates. Therefore the cQASM code and circuit visualization present an arbitrary choice of the gate order. Such ambiguity is eliminated in the LDG. For example, observe that the CP(0,1) gate, which follows T(0), actually commutes with it and therefore the link between them is not present. See reference \cite{Guerreschi2018} for further explanations and examples.}
\label{fig:ldg}
\end{figure}

We briefly discuss how the LDG is built from the list of sequential gates composing the original circuit. The procedure can be described more easily after a few definitions. We refer to two gates as \textit{consecutive} if  they  are  not  separated  by another operation and act on, at least, one common qubit. Moreover we call \textit{parents} (respectively \textit{children}) of gate $g$ all the consecutive and non-commuting gates that logically precede (respectively follow) $g$ in the circuit. The LDG is created by connecting the node representing gate $g$ with an outgoing edge towards all its children
\footnote{For syntactic simplicity and when the situation is clear from the context, we identify the nodes of the LDG with the corresponding gates of the original circuit.}
and with an incoming edge from each of its parents. This can be done by traversing the list of gates sequentially and identifying for each gates its parents (in fact they must be gates that preceded it on the list). The cost of the procedure is at most quadratic in the number of gates, but it is linear when all sets of consecutive and commuting gates have cardinality independent of the total number of gates, the latter being a common situation. The role of commutativity and several examples are discussed in detail in \cite{Guerreschi2018} whereas a simple circuit is discussed in Fig.~\ref{fig:ldg} and its caption.

While the LDG is a machine-independent representation of the algorithm and can undergo dedicated optimization passes on its own, we further assume that one of the preliminary passes has decomposed all the gates (i.e. the LDG nodes). This pass may vary substantially from machine to machine, especially for the quantum computers expected to be built in the short-term period. For example, hardware based on similar superconducting technology may have very different native operations like
partial-iSWAP \cite{Bialczak2010}, 
directed XZ interaction \cite{Corcoles2013}, 
or controlled-phase \cite{DiCarlo2009}. 
Perhaps surprisingly, this situation may become simpler in the long-term since error correction techniques favor discrete gate set composed by only a handful of operations (typically from the $\{H,S,T,CNOT\}$ universal gate set). We consider our scheduler as part of the machine-dependent back-end in the compilation chain.

Following a standard procedure for classical compilers \cite{Cooper1998}, we assign a priority value to each of the vertices of $\L$ or, equivalently, to each of the quantum gates of the algorithm. The first step is assigning a cost to each gate: the cost depends on the figure of merit of the optimization and can be the gate duration, if one wants to minimize the execution time of the algorithm as we will adopt in this work, or the gate infidelity, if one aims at maximizing the fidelity between the experimental run and the ideal circuit \cite{Tannu2019, Nishio2019,Murali2019asplos,Lao2019a}. Then, starting from the vertices without children of the LDG (i.e. the last gates of every logical qubit), one computes the priority of gate $g$ by adding its cost to the highest value among the priorities of its children. This procedure is efficient (see \cite{Guerreschi2018}) and the resulting priority corresponds to a scalar value that can be used to order the gates: a gate can only depend on gates with higher priority and never depends on gates with lower priority. Multiple gates can have the same priority and, in this case, they need to commute whenever they act on at least one common qubit. We denote the priority of gate $g$ with $p(g)$ and observe that it is a machine-dependent quantity since the gate duration is hardware-specific.

\subsection*{OUTPUT: cQASM instructions with explicit time and parallelism}

We observe that the execution of quantum circuits on real hardware requires that the compiler generates micro-instructions that can be directly processed by the electronic controller of the quantum device. Our scheduler produces two kinds of output: The first one is the specification of all electronic signals sent to the quantum hardware (in this work they are given by microwave pules and flux-bias values) at each clock-cycle. The second output is to the interpretation of the micro-instructions in terms of the quantum operations performed at each clock-cycle with explicit parallelism.

The latter output is more suitable for human analysis and is the form we provide in the next sections. However it is important to appreciate that the former kind of output already contains all information on the scheduled circuit and, arguably, in a more fundamental form suitable for direct execution by the electronic apparatus. In section~\ref{sec:hardware} we describe the transmon architecture targeted by the scheduler for the current numerical study, and provide examples of low-level output. 

\begin{figure}[th]
\centering
\includegraphics[scale=0.62]{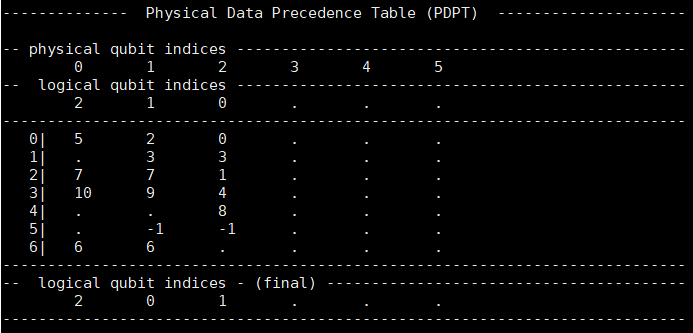}
\caption{The output of our scheduler for the circuit of Fig.~\ref{fig:ldg}. The schedule corresponds to a linear array of 6 qubits with the initial placement $P(0)=2$, $P(1)=1$ and $P(2)=0$ (for a formal definition of the placement function see the next section). Each column corresponds to a physical qubit while each row corresponds to a clock-cycle, with time flowing from top to bottom. A non-negative integer marks the qubit involved in the corresponding gate (see gate index from the LDG in Fig.~\ref{fig:ldg}). Negative integers mark routing operations, here represented by native SWAP gates. 
For this example we assumed that every gate lasts for exactly one clock-cycle. Notice that both SWAP and CP gates are symmetrical and there is no need to specify which qubit is the control one. The last line provides the final qubit placement that takes into account the routing operations and does not need to correspond to the initial placement.}
\label{fig:output}
\end{figure}

Our scheduler generates the output in the form of a table in which each column corresponds to a physical qubit and each row to a clock-cycle. Each entry of the table therefore corresponds to the quantum operation performed on the physical qubit at that given time. We use a non-negative integer to uniquely identify any gate of the circuit, think of it as the index of the corresponding node of the LDG. Since the final schedule includes routing operations that need to be included in the output, we use negative integers to uniquely identify the corresponding gates. For example, if the SWAP gate is available in the target architecture, we may label each routing operation with a negative integer and know that it corresponds to a SWAP gate. In other cases, one needs more gates to decompose a single qubit-exchange operation. An example of the human-readable output is provided in Fig.~\ref{fig:output}


%% file: section_3_snapshot.tex

\section{Execution snapshot and its components}
\label{sec:snapshot}

\extra{Consider naming this section something that indicates you are going to explain the schedule prioritization process. There is a bit of technique explanation as well as definitions and notation.}

We introduce the term execution snapshot, or {\bf snapshot} for short. It refers to the status of the circuit at a given instant of time (or clock-cycle if one expresses the gate duration as a multiple of a common time interval) and includes information on the initial qubit placement, the schedule of the gates already executed, the current qubit placement, and the sub-tree of the LDG with the remaining gates.

Formally, snapshot $\alpha$ is composed by: the pair of functions $P_0$ and $P_\alpha$ corresponding to the initial and current qubit placement respectively, the current schedule provided in either of the formats mentioned in the previous section and denoted with $\sc_\alpha$, and the set $\A_\alpha \subseteq \G$ of those gates already part of the current schedule.
Set $\A_\alpha \subseteq \G$ could be extracted from schedule $\sc_\alpha$, but this may be a non-trivial operation if the schedule is provided in terms of micro-instructions. The schedule itself contains much more information, for example it includes all operations added for the routing (not present in $\A_\alpha$) together with explicit timing and parallelism.

Now we describe the qubit placement function in more detail. Gates in $\G$ act on \textit{logical} qubits, meaning those used in the abstract description of the quantum algorithm. However the machine is composed by \textit{physical} qubits, and the placement is the function describing which physical qubit is associated to which logical qubit. Assume that the algorithm includes $Q_L$ qubits with indices in the range $[0:Q_L-1]$ while the hardware has $Q_P \geq Q_L$ physical qubits with indices in $[0:Q_P-1]$, where $[a:b]$ represents the set of integers from $a$ to $b$, both included.
For {\bf qubit placement} we mean the map that describes which physical qubit is associated to which logical qubit. Specifically, placement $P\,:\,[0:Q_L-1]\rightarrow[0:Q_P-1]$ is the injective function such that $P(j)=k$ means that logical qubit $j$ is associated to physical qubit $k$.

In addition, physical qubits can be seen as the vertices of the {\bf connectivity graph} $\C$, with edges joining pairs of qubits that are connected in the hardware so that an entangling interaction is available. Two-qubit gates are possible only among those qubits. If the physical interaction is asymmetric, leading to asymmetric two-qubit gates, one needs to distinguish between two qubits involved in the gate and this is achieved by having directed edges. In the rest of this work we consider symmetrical interactions leading to symmetric gates (like iSWAP, control-phase, SWAP, \dots) and simply observe that other asymmetric operations, like CNOT gates, can easily be obtained by a local change of basis, \emph{i.e.} by adding a few single-qubit gates.
In addition, we do not include in our treatment the fact that the same kind of quantum operation may be executed with different reliability depending on the specific qubits involved. One may take into account this variability during the scheduling, as suggested in \cite{Tannu2019}.
To summarize, $\C$ is an undirected (unweighted) graph corresponding to the connectivity of the physical qubits.

\begin{figure}[t]
\centering
\includegraphics[scale=0.4]{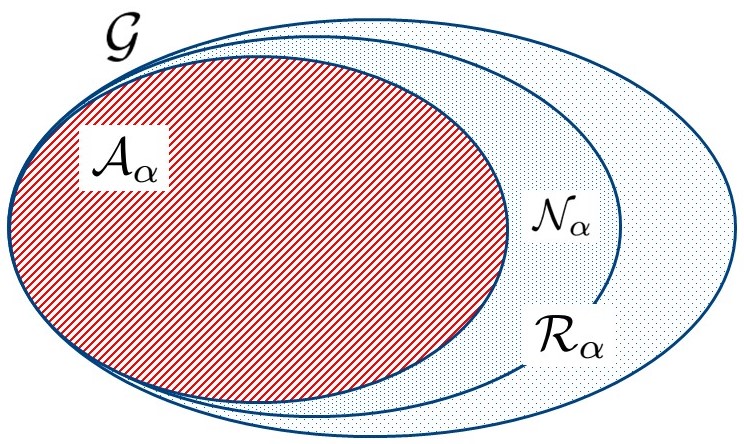}
\caption{Visualization of the sets of quantum gates introduced in the main text. $\G$ is the set of all gates forming the quantum algorithm, whereas the other sets depend on the execution snapshot $\sc_\alpha$. Specifically, $\A_\alpha$ is the set of gates ``already'' scheduled (area with red stripes) and $\R_\alpha$ the set of ``remaining'' gates (area with blue dots that excludes $\A_\alpha$). The latter set contains the subset $\N_\alpha$ of gates that can be scheduled ``next'' (area with denser blue dots) without violating logical dependencies.}
\label{fig:gate_sets}
\end{figure}

Other useful quantities are set $\R_\alpha = \G-\A_\alpha$ containing the gates that remain to be scheduled and set $\N_\alpha \subseteq \R_\alpha$ containing all gates that have no parent in $\R_\alpha$ and that can thus be scheduled next without breaking any logical dependency. The sub-graph of $\L$ obtained by pruning the vertices from $\A_\alpha$ and their edges is denoted $\L_\alpha$. To help navigate the different sets of gates, consider the mnemonic help: $\G$ is the set of all gates, $\A_\alpha$ the set of gates already scheduled, $\R_\alpha$ the set of remaining gates containing the subset $\N_\alpha$ of gates that can be scheduled next. Fig.~\ref{fig:gate_sets} clarifies their relationship.

\begin{figure}[t]
\centering
\includegraphics[scale=0.57]{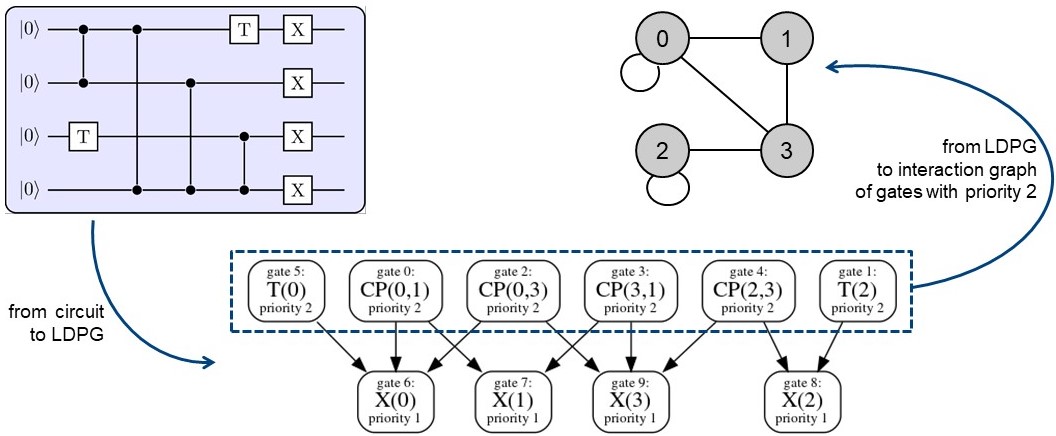}
\caption{Relationship between a quantum circuit, its LDG and the interaction graph of a specific snapshot. For this example, we consider the initial snapshot in which all gates still remain to be scheduled and, therefore,
the set of gates $\N_\alpha$ (i.e. those gates to be scheduled next) can be identified with the set of gates with priority 2. The interaction graph $\I_\alpha$ reflects this situation: the logical qubits correspond to the vertices, while one-qubit gates are represented by self-edges and two-qubit gates by undirected edges. It is important to observe that the interaction graph is not related to the hardware connectivity. In the following, to avoid any confusion between the two objects, we will use circular shapes to depict logical qubits in the interaction graph, but will adopt hexagonal shapes to depict physical qubits in the connectivity graph.}
\label{fig:interaction}
\end{figure}

Finally, to each set $\N_\alpha$ can be associated an undirected graph, called {\bf interaction graph} in \cite{Guerreschi2018} and here denoted by $\I_\alpha$. Each vertex of the interaction graph represents a logical qubits (for a total of $Q_L$ vertices) and each edge corresponds to a gate from $\N_\alpha$: for example a one-qubit gate on logical qubit $j$ is represented by a self-edge on vertex $j$, while a (symmetric) two-qubit gate involving qubits $j$ and $k$ is represented by an undirected edge connecting vertices $j$ and $k$. Fig.~\ref{fig:interaction} shows the interaction graph for a typical quantum circuit where set $\N_\alpha$ is identified with the set of gates with priority 2 (here being pairwise commuting gates CP and T).

%% file: section_4_protocol.tex

\section{Scheduling protocol and division in sub-routines}
\label{sec:protocol}

Any quantum circuit can be scheduled by selecting the operations to execute next given the current snapshot. The strategy guiding the decision process must enforce that the schedule is correct, but it also determines the quality of the schedule in terms of the total number of operation or clock-cycles required to execute the complete circuit. The fewer gates or clock-cycles, the more optimized is the schedule.

Assume the execution is currently at the snapshot $\alpha$. The next gates to schedule are contained in $\mathcal{N}_\alpha$ and we observe that if two gates $g_1,g_2 \in \mathcal{N}_\alpha$ act on at least one common qubit then they must commute. Also recall that any gate $g \in \mathcal{N}_\alpha$ acts on at most two qubits.

\subsection{Scheduling protocol}

\extra{Consider format the protocol with ``algorithmicx''.}

\begin{figure}[t]
\centering
\includegraphics[scale=0.53]{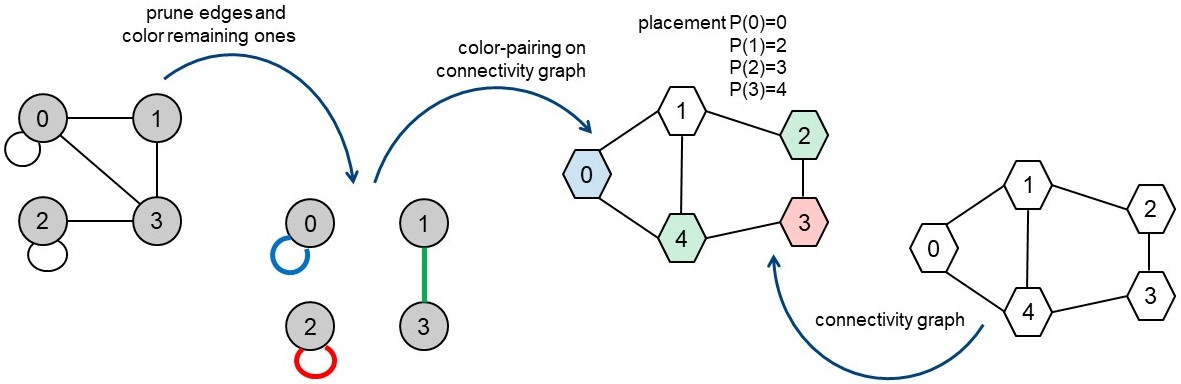}
\caption{
\textbf{Leftmost insert:} the interaction graph $\I_\alpha$ from Fig.~\ref{fig:interaction}.
\textbf{Second insert from left:} prued interaction graph $\I$ in which every vertex has at most one edge.
\textbf{Rightmost insert:} connectivity graph $\C$ of the physical qubits, here depicted as hexagons to visually distinguish them from the circular-shaped logical qubits. 
\textbf{Second insert from right:} instance of the color-pairing problem that is defined using both $\I$ and $\C$, together with the qubit placement function $P$ explicitly indicated in figure.}
\label{fig:color-pairing}
\end{figure}

The protocol consists in the following steps:
\begin{enumerate}
\item[A.] Create the interaction graph $\I_\alpha$ of the current snapshot by adding one edge for each gate in $\mathcal{N}_\alpha$. It may have vertices with more than one edge. Eliminate edges until every vertex has at most one edge (with self-edge counting one), call this graph $\I$.
\item[B.] Prune the interaction graph $\I_\alpha$ by eliminating one edge at a time until every vertex has at most one edge (with self-edge counting one), call this graph $\I$.
\item[C.] Color each edge of $\I$ with a different color. Observe that the current placement function $P_\alpha$ can be used to associate to each vertex of $\I$ (i.e. a logical qubit) a vertex of the connectivity graph $\C$ (i.e. a physical qubit). Transfer the color from the edges of $\I$ to the vertices of $\C$: formally, if edge $(j,k)$ of $\I$ is red, color vertices $P_\alpha(j)$ and $P_\alpha(k)$ of $\C$ in red. See Fig.~\ref{fig:color-pairing} for a visual definition of this step and the previous one.
\item[D.] Solve the color-pairing problem for graph $\C$ as defined below \cite{Guerreschi2018}: the goal is to achieve a color pattern such that all pairs of vertices with the same color are connected by an edge. To achieve a compatible color pattern, one has a single move available: the exchange of colors of two vertices connected by an edge. In the context of this work, this operation is an abstraction of the effect of a SWAP gate on the placement function $P_\alpha$. Consequently, $P_\alpha$ will be updated by the color-pairing solver. The goal of color-pairing corresponds to achieve a placement in which all gates from $\I$ (i.e. its edges) are between connected physical qubits.
\item[E.] Before the color-pairing problem is fully solved one reaches the situation where only a subset of gates is between connected physical qubits and one can already schedule the compatible gates for execution. Once the gate is scheduled, the corresponding edge of $\I$ can be removed and the vertices of $\C$ representing the involved qubits become colorless. One can dynamically add more edges from $\I_\alpha$ to $\I$ (and then color the vertices of $\C$) choosing among the edges that were eliminated in step B.
\item[F.]  The previous step provides limited advantage when all (or most) edges of $\I_\alpha$ have already been scheduled for execution or are in the current graph $\I$. Even before the complete solution of the color-paring problem in step D, one can therefore update the execution snapshot to the new snapshot $\alpha^\prime$ and restart from step A.
\end{enumerate}

The sequence of steps A-F has to be repeated until all gates in $\G$ are scheduled, meaning that both $\N_\alpha$ and $\R_\alpha$ are empty. Fig.~\ref{fig:color-pairing} provides a visualization of steps A-C leading to the definition of the color-pairing instance.
Most importantly, the above procedure describes the scope of each steps more than the precise way in which they are carried out. To specify the actual protocol, we break it down in three separate tasks that we will address with specific strategies. In the next sub-sections we present a few policies for these three tasks.

\subsection{Division in sub-routines}

While discussing the scheduling protocol we avoided to address how the initial placement of the qubits, $P_0$, was determined. In practice, the strategy to initialize the qubit placement (or logical-to-physical qubit-index map) can be identified as a separate sub-routine. Including it in the list, the sub-routines that needs to be addressed to make the above protocol concrete are:

\vspace{3mm}
\begin{description}
\item[SR0: initial placement of the qubits] Associate each logical qubit to a specific physical qubit before scheduling first gate that involvs it.
\item[SR1: pruning the interaction graph] Select which edges of the interaction graph $\I_\alpha$ are eliminated to reach $\I$. In general we want to preserve as many edges as possible, compatibly with the constraints that every vertex has at most one connected edge. This is related to steps (1)-(2) of the procedure.
\item[SR2: solving color-pairing problem] Provide a sequence of color exchanges (the abstract representation of SWAP operations) that achieves a compatible color pattern. The effective strategy should be able to deal with dynamical color-pairing instances in which the same-color vertices become colorless once they are connected by an edge and new colors may appear in colorless vertices, according to step (5). 
\item[SR3: updating the snapshot] Provide a list of criteria to reset the procedure back to step (1) before all gates of $\N_\alpha$ have been scheduled. Intuitively, the more often we update the snapshot the larger number of opportunities of optimization we expose to SR1 and SR2.
\end{description}

\extra{Add discussion and performance for each choice: Which steps are more important than others (if any)? What is the relative performance gain that happens as a result of each stage? Within a stage, how do alternative implementations compare to one another?}

\subsection*{SR0: Initial placement of the qubits}

\begin{description}
\item[trivial]
Logical qubit $j\in[0:Q_L-1]$ is associated with physical qubit $j$. If $Q_P>Q_L$, the remaining physical qubits are not associated with any logical one but can still be involved in routing operations.
\item[random]
Logical qubits are associated uniformly at random to the physical qubits. 
%
%
\item[subgraph]
Consider the connectivity graph $\C$ and the interaction graph $\I_0$ of the very first execution snapshot $\alpha=0$. Find the maximum common edge subgraph between $\C$ and $\I_0$ and use the corresponding vertex homomorphism to define the original placement $P_0$. Since the maximum common edge subgraph is a NP hard problem, we implemented an heuristic based on breadth-first-search to find approximate solutions. Our heuristic is deterministic, but for stochastic alternatives evaluating several schedules starting from different $P_0$ may be beneficial to the overall scheduling task.
\end{description}

\vspace{2mm}
The trivial policy has been included to allow the user to impose a specific placement by changing the input file. The random policy represents the non-deterministic baseline. The subgraph policy is expected to be superior to the baseline when the number of possible initial placements is too large. This already happens for a small number of logical qubits even in physical registers of minimal size (with $Q_P=Q_L$) since there is a combinatorial number of permutations of $Q_L$ objects. We quantify this situation in Section~\ref{sec:numerical-evaluation}. The heuristic used for the maximum common edge subgraph problem is described in Appendix~\ref{app:sec:subgraph-heuristic}.

\subsection*{SR1: Pruning the interaction graph}

\begin{description}
\item[random]
Randomly select an edge in $\I_\alpha$ and eliminate the other edges that are connected to the same vertices. Continue with random edge selection until all conflicts are eliminated. Recall that the resulting graph is denoted with $\I$.
\item[one-qubit-first]
As above, but consider the self-edges of $\I_\alpha$ first. They corresponds to one-qubit gates.
\item[lowest-index-first ]
Starting from the edge in $\I_\alpha$ corresponding to the gate that occurred earliest in the original list of gates of the algorithm, eliminate the other edges that are connected to the same vertices. Continue with the gate occurring next in the original list until all conflicts are eliminated.
\item[edge-coloring]
Solve the edge-coloring problem for $\I_\alpha$ in a non-deterministic way by using greedy search. If more than one color is required, choose the one coloring most edges and eliminate all edges of different color. The remaining graph is $\I$.
\end{description}

The last policy has not been implemented in this work, but it is the natural refinement of the random approach when the edge coloring is solved in a stochastic way. In the context of one-dimensional qubit arrays it has been discussed and implemented in \cite{Guerreschi2018}.

\subsection*{SR2: solving color-pairing problem}

\begin{description}
\item[left-accumulation]
Specialized to one-dimensional array of qubits. Starting from the left edge of the array, if a colored vertex is not adjacent to the same-color vertex, move the latter towards the left edge until the pair is connected. Continue scanning the array until the right edge is reached. In \cite{Guerreschi2018} it was demonstrated that this policy requires the smallest number of color exchanges (i.e. SWAP gates) to achieves a compatible color pattern. However, while the ``left accumulation'' minimizes the number of SWAPs, it is easy to find an example in which it does not minimize circuit depth.
\item[dynamical-pattern-improvement]
One of the main contributions of this work is the proposal of a strategy to solve the color-pairing problem that is based on dynamical pattern improvement. This strategy is described in the next section.
%
\end{description}


\subsection*{SR3: updating the snapshot}

\begin{description}
\item[no more next gates]
Do not update the snapshot until all gates from $N_\alpha$
have been scheduled, corresponding to the complete schedule of $\I_\alpha$ (and not only $\I$). When updating the snapshot $\sc_\beta$, only the gates with largest priority (possibly different for each logical qubit) are used to populate the new set $N_\beta$.
%
%
\item[always] As above, but the snapshot update takes place as soon as a gate is scheduled. This policy increases the chance of parallelism and optimization, especially when associated to the SR2 policy of dynamical pattern improvement. 
\item[always despite priority]
As above, but one accepts gates in $\N_\alpha$ that acts on a common qubit but have different priority. In section~\ref{sec:numerical-evaluation} we will discuss why this limitation is important to schedule certain highly structured quantum circuits.
\end{description}


%% file: section_5_dynamical-pattern-improvement.tex

\section{Dynamical pattern improvement}
\label{sec:dynamical-pattern-improvement}


The primary contribution is a novel approach to solving SP2, a solution we call dynamical pattern improvement. Consider the color-pairing problem: it is always possible to quantify how far from a compatible pattern the current one is. For each color, find the shortest path in $\C$ between the two vertices and subtract 1: this corresponds to the minimum number of color exchanges needed to move the color to connected vertices (if a single vertices has that color, such value is 0). Sum these contributions from all colors to obtain the color-pairing distance of the current pattern. We observe that a color exchange affects at most two terms of the color-pairing distance, which can therefore changes only by $\delta\in\{+2,+1,0,-1,-2\}$, with $\delta=-2$ being the best case.

Consider that the current snapshot is $\alpha$, determine $\N_\alpha$ and create the corresponding interaction graph $\I_\alpha$. The latter in general contains both self-edges and regular edges, and can be pruned according to the SR1 policy ``one-qubit first''.

We apply moves that 1) can be performed in parallel and 2) tends to improve the pattern maximally. Notice that each self-edge of $\I$ corresponds to a vertex of $\C$ that has a unique color. Schedule the corresponding (single-qubit) gates. Analogously, schedule all two-qubit gates corresponding to connected same-color vertices. Loop on the edges of $\C$ (excluding those connected to the vertices just considered) and exchange the color of two connected vertices when it is favorable to reduce the color-pairing distance by 2. Then loop again on the edges and perform the exchanges that reduce the color-pairing distance by 1. Finally, loop once more and perform the exchanges that leave the color-pairing distance unaltered%
, but that somehow improve the pattern according to the pseudo-code provided in the next subsection.

To provide as many opportunities as possible for color exchanges that reduce the color-pairing distance by 2, we update the snapshot with always SR3 policy.

\subsection*{Pseudo-code for dynamical pattern improvement}

Here we present pseudo-code to implement the policy for SR2 described in the previous section. Recall that each vertex color of $\C$ corresponds to a gate from $\N_\alpha$, for this reason we use $g_i$ to indicate the quantum gate, the corresponding edge of $\I$ and the vertex-color of $\C$. Denote with $V_\C=[0:Q_P-1]$ and $E_\C \subseteq \N_\alpha$ the set of vertices and edges of $\C$ respectively.

Consider vertex $j\in V_\C$ with color $g$, if gate $g$ corresponds to a two-qubit gate then there is another vertex $k\in V_\C$ with the same color $g$. Denote with $d(j,k)$ the length of the shortest-path from vertex $j$ to $k$ (recall that one needs at least $d(j,k)-1$ color exchanges to have the new vertices with color $g$ connected). By convention $d(j,j)=0$, valid for one-qubit gates.

In the pseudocode we use a few intuitive functions:
\begin{itemize}
\item $\textsc{Color}(j)$ returns the color of vertex $j\in V_\C$;
\item $\textsc{Qubits}(g)$ returns the two logical qubits on which gate $g$ acts. If $g$ is a one-qubit gate, it returns twice the same qubit;
\item $\textsc{Schedule}(g)$ returns a boolean indicating whether gate $g$ can be scheduled avoiding conflicts with control settings for previously scheduled operations. When returning $\text{True}$, it also schedules gate $g$;
\item $\textsc{PruneEdgesOfVertex}(j,\C)$ eliminates from $\C$ the edges connected to vertex $j$;
\item $\textsc{ColorExchange}(e,P,\C)$ returns a boolean indicating whether a SWAP operation along the edge $e=(i,j)$ of $\C$ can be scheduled avoiding conflicts with control settings for previously scheduled operations. When returning $\text{True}$, it also schedules the corresponding SWAP gate, exchange the colors of $i,j$ and update the qubit placement function $P$.
\end{itemize}



\begin{algorithm}[H]
\caption{Dynamical pattern improvement of color-pairing}
\label{alg:pattern-improvement}
\begin{algorithmic}[1] 
	\Procedure{PatternUpdate}{$\C$}
	\Comment{$\C$ is connectivity graph with colored vertices}
		\State $P() \gets P_\alpha()$
		\Comment{placement function}
		\State $pass \gets 0$
		\While {$pass<5$}
			\ForAll {$e=(j,k) \in E_\C$}
			\Comment{avoiding edges already pruned}
				\State $g \gets \Call{Color}{j}$
				\State $q_1,q_2 \gets \Call{Qubits}{g}$
				\If {$P(q_2) = j$}
				\Comment {order $q_1,q_2$ so that $P(q_1)=j$}
					\State $q_1,q_2 \gets q_2,q_1$
				\EndIf
				\State $g^\prime \gets \Call{Color}{k}$
				\State $q_1^\prime,q_2^\prime \gets \Call{Qubits}{g^\prime}$
				\If {$P(q_2) = k$}
				\Comment {order $q_1^\prime,q_2^\prime$ so that $P(q_1^\prime)=k$}
					\State $q_1^\prime,q_2^\prime \gets q_2^\prime,q_1^\prime$
				\EndIf
				\State $\delta \gets d(k,P(q_2))+d(j,P(q_2^\prime))-d(j,P(q_2))-d(k,P(q_2^\prime))$
				\Switch{$pass$}
					\Case{0}
					\Comment{schedule 1-qubit gates}
						\If {$q_1 = q_2$}
							\If {$\Call{Schedule}{g} = \text{True}$}
								\State \Call{PruneEdgesOfVertex}{$j,\C$}
							\EndIf
						\EndIf
						\If {$q_1^\prime = q_2^\prime$}
							\If {$\Call{Schedule}{g^\prime} = \text{True}$}
								\State \Call{PruneEdgesOfVertex}{$k,\C$}
							\EndIf
						\EndIf
					\EndCase
					\Case{1}
					\Comment{schedule compatible 2-qubit gates}
						\If {$g = g^\prime$}
							\State \Call{Schedule}{$g$}
							\Comment {even if $g$ cannot be scheduled, no color exchange for $j,k$}
							\State \Call{PruneEdgesOfVertex}{$j,\C$}
							\State \Call{PruneEdgesOfVertex}{$k,\C$}
						\EndIf
					\EndCase
					\Case{2}
					\Comment{schedule color-exchange with $\delta=-2$}
						\If {$\delta=-2$}
							\If {$\Call{ColorExchange}{e,P,\C}=\text{True}$}
								\Comment {if successful, placement $P$ changes}
								\State \Call{PruneEdgesOfVertex}{$j,\C$}
								\State \Call{PruneEdgesOfVertex}{$k,\C$}
							\EndIf
						\EndIf
						\If {$\delta>0$}
						\Comment{bad color-exchange should not be considered anymore}
							\State \Call{PruneEdge}{$e,\C$}
						\EndIf
					\EndCase
					\Case{3}
					\Comment{schedule color-exchange with $\delta=-1$}
						\If {$\delta=-1$}
							\If {$\Call{ColorExchange}{e,P,\C}=\text{True}$}
								\State \Call{PruneEdgesOfVertex}{$j,\C$}
								\State \Call{PruneEdgesOfVertex}{$k,\C$}										\EndIf
						\EndIf
					\EndCase
					\Case{4}
					\Comment{this imply that $\delta=0$}
						\State $r \gets \Call{UniformRandom}{0,1}$
						\If {$r<t_2/t_\text{SWAP}$}
							\If {$\Call{ColorExchange}{e,P,\C}=\text{True}$}
								\State \Call{PruneEdgesOfVertex}{$j,\C$}
								\State \Call{PruneEdgesOfVertex}{$k,\C$}										\EndIf
						\EndIf
					\EndCase
				\EndSwitch
			\EndFor
			\State $pass \gets pass+1$
		\EndWhile
	\EndProcedure
\end{algorithmic}
\end{algorithm}

\subsection*{Algorithmic analysis}

The algorithm for the dynamical pattern improvement of color-pairing is provided as Algorithm~\ref{alg:pattern-improvement}. It consists of 5 passes to schedule gates that contribute to the color-pairing solution, in order from what we consider the most to the least advantageous. The most effective way to perform routing is to schedule as many color exchanges as possible with $\delta=-2$. Notice that this is possible only when both colors involved in the exchange corresponds to two-qubit gates. For this reason we want as many two-qubit gates as possible among $E_\C$.

This is why in the first pass we schedule all possible one-qubit gates: this allows us to consider more two-qubit gates in the next iteration of \textsc{PatternUpdate} (recall that, according to the policy for task C, we update the snapshot after each iteration of \textsc{PatternUpdate}).

In the second pass we schedule the two-qubit gates from $E_\C$ that are already between physically connected qubits. if only the connectivity was considered among the physical constraints, pass 1 and 2 could be combined. This is not possible when control settings for two-qubit gates may interfere with one-qubit gates on neighbor qubits. For example, the concrete architecture discussed in section~\ref{sec:hardware} does not always allow one to combine the first two passes.

From now on, the only operations we possibly schedule are SWAP gates corresponding to the exchange of vertex-colors. Notice that the quantum architecture may not provide the SWAP gate among the physically available gates: if this is the case, we use SWAP gate as a reference to the suitable decomposition of a SWAP operation. before scheduling a routing operation, we always verify that it does not conflict with previously scheduled gates.

In the third pass we schedule those SWAPs that reduce the overall color-pairing distance by $\delta=-2$, while in the fourth pass we schedule the SWAPs with $\delta=-1$.

Finally, in the fifth pass we schedule the SWAPs with $\delta=0$ in a non-deterministic way: since it is unclear whether the color exchange is actually favorable, we perform it with probability given by the ratio $t_2 / t_{SWAP}$, where $t_2$ is the duration of a native 2-qubit operation of the hardware and $t_{SWAP}$ is the duration of the SWAP operation. An alternative choice would have been to consider the priority of the gates involved and apply the color exchange only if it benefits the gate with higher priority of the pair.

Notice that during the procedure graph $\C$ is modified, in particular some of its edges from $E_\C$ are pruned, and therefore we assume that a working copy is actually received as input and irreversibly modified for operational convenience. The original connectivity graph, not pruned, remains stored somewhere else.

A frequent operation is, given two vertices $j,k\in V_\C$ with the same color $g$, to compute the distance $d(j,k)$. To avoid recomputing $d(j,k)$ very often and therefore to reduce the computational cost of the algorithm, the values $d(j,k)$ can be stored in an array: the memory required to store all distances is $\mathcal{O}(Q_P^2)$.


%% file: section_6_hardware.tex

\section{Hardware description and control electronics}
\label{sec:hardware}

In this work we provide a general procedure to schedule quantum circuits on quantum machines with restricted connectivity that can accommodate additional constraints from the control electronics while reducing the routing overhead. However any quantitative evaluation of the efficacy of the suggested policies must fully specify the target hardware for the compilation. For this reason, here we describe the transmon architecture that we consider in the numerical part of our work.

While the connectivity graph $\C$ may not reflect that of actual chips, both the kind of superconducting qubits and their operability are taken from the architecture developed in DiCarlo's group at QuTech and TUDelft together with Intel. For simplicity, we will refer to the architecture of this section simply by ``transmon architecture'' or ``this'' architecture. An extensive description is provided in reference \cite{Versluis2017}, here summarized with the scope of clarifying its features that are relevant to the compilation task.

Every physical qubit has a fundamental frequency corresponding to the energy space between its computational levels $\ket{0}$ and $\ket{1}$. More than one qubit may have the same frequency. The qubits can be controlled with signal at (or near) their fundamental frequency.

Single-qubit gates are implemented via microwave pulses, we assume that each physical qubit has a dedicated drive line from an Analog Waveform Generator (AWG). All AWGs have a finite number of waveforms stored in memory, each corresponding to a specific one-qubit gates. We assume that all AWGs have the ability to perform the same set of gates. Pulses generating the same quantum gate (but on different qubits) have the same duration, typically a small integer multiple of a fix clock-cycle duration. Notice that if multiple drive lines were connected to the output of the same AWG, the parallelism may be reduced depending on the spectral separation of the various pulses.

Two-qubit gates are implemented in a passive way: transmon qubits are formed by superconducting circuits that can be connected with resonators. If the connected qubits have the same frequency then they interact, otherwise if their detuning is larger than a certain threshold then their interaction can be neglected. In the machine, the resonators connect qubits according to the connectivity graph $\C$ and the fundamental qubit frequency are chosen in such way that all interactions are suppressed. Through flux-bias lines one can tune the resonant frequency of each qubit independently, in this way created resonances between connected qubits and effectively switching on their interaction. With appropriate timing, the interaction results in a controlled-phase gate.

With the flux-bias, each qubit can be set to one of three frequencies: its fundamental frequency $f_a$, the interaction frequency $f_a^\text{int}$ typically corresponding to a fundamental frequency of some qubit connected with it, and the parking frequency $f_a^\text{park}$ intermediate between the two and used for the readout or when unwanted resonances need to be avoided. The subscript $a$ uniquely labels each set of three frequencies. For convenience, we will say that a qubit has frequency $f_a$ and three possible ``tunes'': $f_a$, $f_a^\text{park}$, $f_a^\text{int}$.

\begin{figure}[t]
\centering
\includegraphics[scale=0.5]{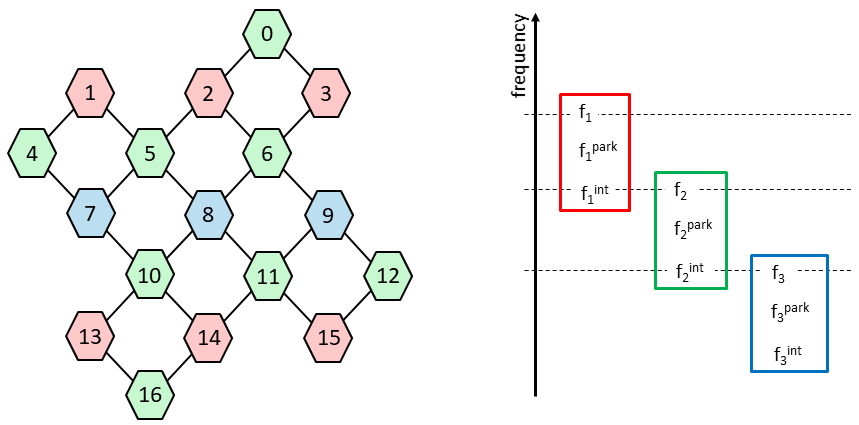}
\caption{
\textbf{Left panel:} Qubit register with connectivity and qubit labels. In the implementation of surface code, red and blue hexagons correspond to data qubits, while green hexagons correspond to ancilla qubits for X-type and Z-type stabilizer measurements respectively. The lines connect pairs of qubits that are available for controlled-phase operations. These are the only two-qubit gates available in the architecture and are implemented by bringing the connected qubits in resonance and mediating the interaction with passive resonators.
\textbf{Right panel:} Frequency scheme for the qubits. All ancilla qubits have the same fundamental frequency $f_2$, intermediate between the high frequency $f_1$ of red qubits and the low frequency $f_3$ of the blue qubits.}
\label{fig:machine}
\end{figure}

Fig.~\ref{fig:machine} illustrates a bidimensional array of qubits with the connectivity of a partial square grid. The color code of each qubit indicates their fundamental frequency (high frequency $f_1$ for the red qubits, medium frequency $f_2$ for the blue and green ones, and low frequency $f_3$ for the pink qubits). Both panels are adapted from reference \cite{Versluis2017}.

It is worth mentioning a few ways in which the gate parallelization is affected by the chip architecture and control electronics. For two-qubit gates, one needs to avoid unwanted resonances and this means that, considering the layout in Fig.~\ref{fig:machine} in which qubit 0 is connected to qubit 2 and 3 having the same frequency, the execution of a gate between 0 and 2 requires them to be in resonance and, therefore, qubit 3 has to be forced out of resonance, possibly by changing its tune to $f_2^\text{park}$.

Concerning one-qubit gates, multiple qubits can be connected to the same line: this implies that at most one kind of microwave pulse can be sent per distinct qubit frequency (and this only if frequency multiplexing and parallelism is enabled). Therefore one cannot perform two different one-qubit operations on two qubits connected to the same drive line if they have the same frequency too. The possibility of performing the same gate on multiple qubit with the same frequency may be possible if spatial parallelism is enabled.

In our numerical experiments of next section, we vary the number of frequencies in the chip discuss results suggesting that going beyond a handful of distinct frequencies provide an increasingly small parallelization gain.


%% file: section_7_numerical-evaluation.tex

\section{Numerical evaluation}
\label{sec:numerical-evaluation}

The chip design in reference~\cite{Versluis2017} was explicitly designed to be a modular patch for surface-code implementations with arbitrarily large code distance. Nonetheless the constraints are well defined for arbitrary qubit connectivity once we specify the fundamental frequency of each physical qubit and the group of qubits connected to the same drive lines from the analog waveform generator.

In the numerical evaluation we consider two hardware topologies, a one-dimensional array of qubits and a two-dimensional grid. The quantum circuits to test the different policies we introduced in Section~\ref{sec:protocol} are from two classes: Quantum Approximate Optimization Algorithm (QAOA) or Quantum Fourier Transform (QFT).


\subsection{Quantum Approximate Optimization Algorithm}
\label{subsec:qaoa}

\extra{What point are you going to convey in the numerical analysis? I think you currently do a good job of describing this benchmark and an application of your compiler operating on it, with a specific set of policies. Do you have any results breaking down the significance of each individual policy?}

The Quantum Approximate Optimization Algorithm \cite{Farhi2014} emerged in recent years as a strong candidate for practical applications on devices without error correction. It is a variational algorithm in which the quantum circuit is described in terms of a few classical parameters $\gamma$ (typically the angle of rotation for one- and two-qubit gates) whose values are iteratively improved by a classical optimization procedure. The goal is to find the parameter values for which the state at the end of the quantum circuit is as close as possible to the solution of the combinatorial problems at hand.

For the scope of this study, we only need to describe the specific QAOA circuits used to benchmark our scheduler and compare its different policies. In practice, the circuits of this section are used to solve a graph partitioning problem called MaxCut: the goal is to assign each node of the graph to either of two partitions in such a way that the maximum possible number of edges connect vertices of different partitions. MaxCut is an NP-hard problem and has application in machine scheduling \cite{Alidaee1994}, image recognition \cite{Neven2008}, electronic circuit layout \cite{DEZA1994191}, and software verification$/$validation \cite{Jose2011, Guo2013}.

To understand the task of scheduling QAOA circuits, we observe that they are composed by a sequence of $p$ layers, each layer being composed by: 1) two-qubit ZZ rotations, one for each and every edge of the graph to partition and all with the same rotation angle, and 2) one-qubit X rotations, one for each and every qubit and all with the same rotation angle.

\begin{figure}[h]
\centering
\includegraphics[scale=0.5]{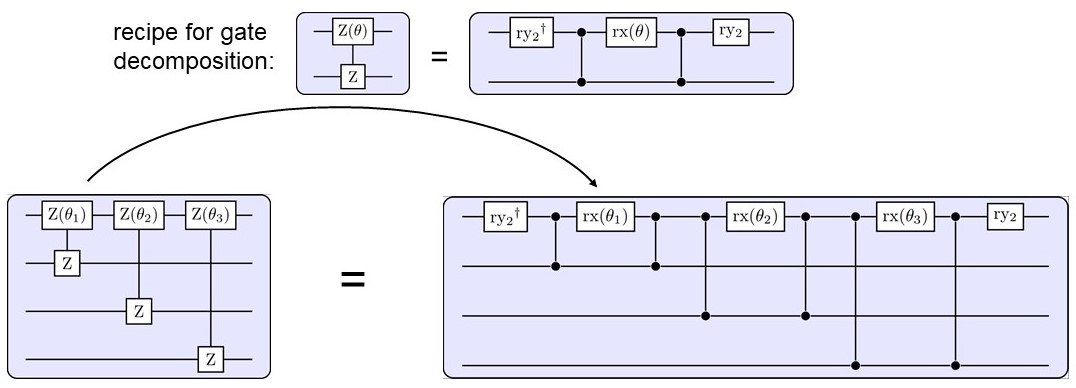}
\caption{Gate decomposition of the two-qubit ZZ rotations in the transmon architecture of \cite{Versluis2017}.The top panel shows the decomposition of a single ZZ rotation while the bottom panel shows a trivial simplification when multiple ZZ rotations involve a common qubit. Notice, however, that the simplification would have not been possible if the single qubit rotations were not applied to the same qubit, here the topmost one. In general, the decomposition imposes a specific order to the ZZ rotations and limits the optimization opportunities for the scheduler. The development and implementation of more advanced gate simplifications is beyond the scope of this work.}
\label{fig:qaoa-gates}
\end{figure}

While the ZZ interaction is the native interaction for the transmon architecture we consider, it is used to implement a non-parametric two-qubit gate, the controlled phase (CP): its only non-trivial effect is adding a minus sign (i.e. a $\pi$ phase) when both qubits are in state $\ket{1}$. To implement the ZZ rotation, we apply the decomposition in Fig.~\ref{fig:qaoa-gates}. However, this  strongly limits the possibility of optimization since multiple commuting ZZ rotations becomes not commuting once decomposed: to explore how the scheduler takes advantage of the gate commutation, we also assume a modified architectures where direct ZZ rotations are allowed.

\begin{figure}[t]
\centering
\includegraphics[scale=0.45]{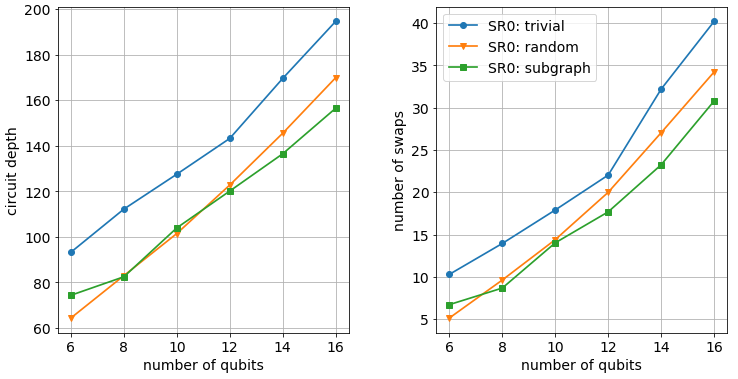}
\caption{Impact of the policy for initial qubit placement (SR0) on the schedule of QAOA circuits with $p=1$. Schedules obtained for the 17-qubit chip described in \cite{Versluis2017}) after appropriate gate decomposition. Each point represents the average of 224 instances of 3-regular graphs, and each circuit has been scheduled 2000 times. The statistical fluctuations of each point are of the same magnitude as in Fig.~\ref{fig:qaoa-scaling}, but are not included here to increase the readability. The other policies are: SR1 one-qubit-first, SR2 dynamical-pattern-improvement, SR3 always.}
\label{fig:qaoa-placement}
\end{figure}

First of all, we consider the unmodified trasmon architecture that requires gate decomposition for ZZ rotations. We aim at quantifying the impact of the initial qubit placement and therefore schedule the shallowest QAOA circuits composed by a single layer, i.e. for $p=1$. The results are shown in Fig.~\ref{fig:qaoa-placement} in which we average 224 instances of MaxCut on 3-regular graphs. Each circuit has been scheduled 2000 times and we report the instance average of the best circuit schedule in terms of circuit depth and number of additional swaps. One observes that for small number of qubits $Q_L$ the random placement performs better than the subgraph policy. This is due to relatively lucky placements that improve over the deterministic one obtained with the subgraph policy. When the qubit number increases, however, the lucky events become harder to find and the subgraph policy performs best.

\begin{figure}[b]
\centering
\includegraphics[scale=0.45]{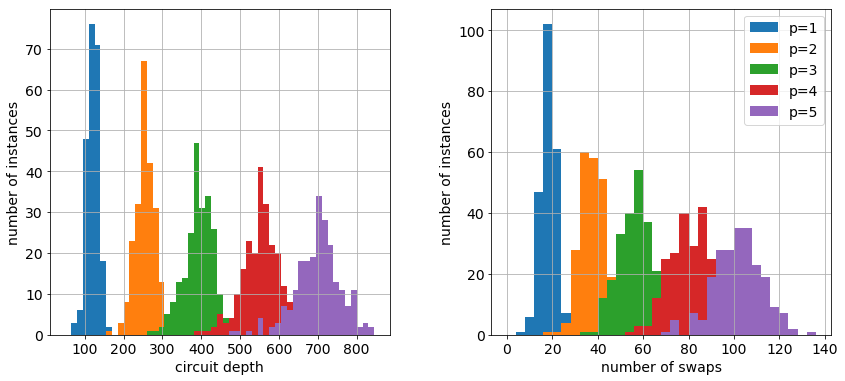}
\caption{Histograms dividing the QAOA instances depending on the circuit depth or  number of SWAPs of the minimal schedule among 2000 trials. Each instance has $Q_L=12$ qubits and is scheduled for the 17-qubit transmon chip from \cite{Versluis2017}. The policies are: SR0 subgraph, SR1 one-qubit-first, SR2 dynamical-pattern-improvement, SR3 always.}
\label{fig:qaoa-fluctuations}
\end{figure}

\begin{figure}[t]
\centering
\includegraphics[scale=0.45]{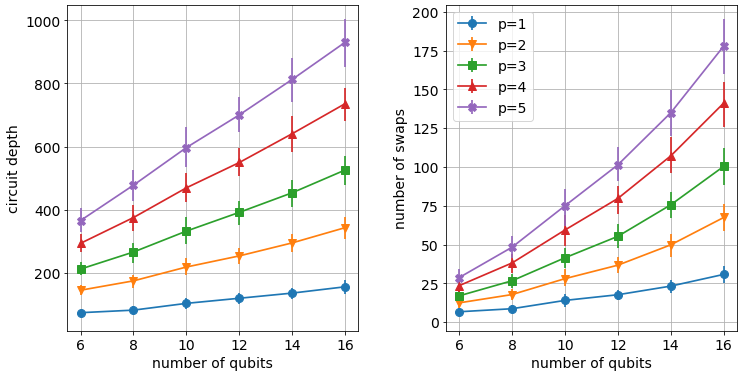}
\caption{Circuit depth as a function of the number of qubits $Q_L$ for different numbers of QAOA layers. Schedules obtained for the 17-qubit chip described in \cite{Versluis2017}). Each point represents the average of 224 instances of 3-regular graphs. The policies are: SR0 subgraph, SR1 one-qubit-first, SR2 dynamical-pattern-improvement, SR3 always.}
\label{fig:qaoa-scaling}
\end{figure}

For the remaining tests, we adopt the placement policy SR0 subgraph. We investigate how does the circuit depth and number of additional SWAP gates scales with the number of logical qubits $Q_L$ of QAOA circuits. We consider 224 instances of MaxCut for random 3-regular graphs with variable number of vertices from 6 to 16. We schedule the circuits for execution in the 17-qubit chip from \cite{Versluis2017}. In Fig.\ref{fig:qaoa-fluctuations} we quantify the statistical fluctuations due to the stochastic nature of the MaxCut instances, while the scaling results are presented in Fig.~\ref{fig:qaoa-scaling}. Here the error bars indicate the 15\% and 85\% percentile and the different curves are characterized by a different number $p$ of QAOA layers.

One surprising result is that, at fixed number of qubits, the number of SWAP gates increases more than linearly compared to the $p=1$ case. This is an undesirable situation since one can always insert the routing operations used for $p=1$ in reverse order and implement the second layer of QAOA operations and so on. Since the decomposed ZZ rotations do not always commute, the inverse routing strategy is possible only when we invert the order of ZZ rotations for the layers with even index. We considered this expedient when computing the results of Fig.~\ref{fig:qaoa-scaling}. Despite this precaution, the optimized schedules do not correspond to sublinear scaling in our numerical study.

On the positive side, the comparatively better schedule for $p=1$ can be straightforwardly applied to longer QAOA circuits by run it in reverse and making the number of swaps and depth strictly linear in $p$. The compact schedule at $p=1$ may be a further indication of the efficacy of the initial qubit placement, the policy SR0 subgraph.

\begin{figure}[b]
\centering
\includegraphics[scale=0.45]{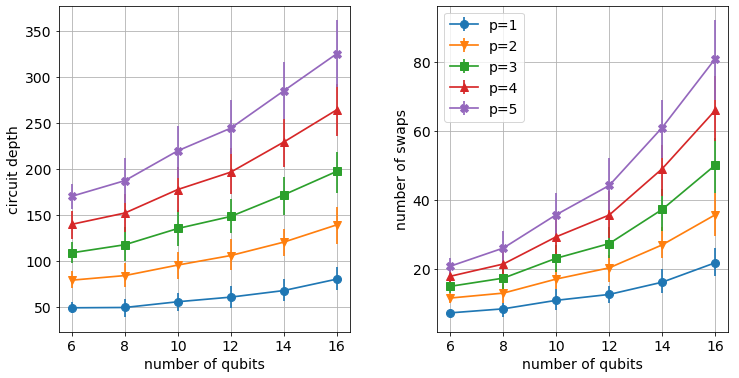}
\caption{Circuit depth as a function of the number of qubits $Q_L$ for different numbers of QAOA layers. Schedules obtained for a modified transmon architecture with native gate set including ZZ rotations. Each point represents the average of 224 instances of 3-regular graphs. The policies are: SR0 subgraph, SR1 one-qubit-first, SR2 dynamical-pattern-improvement, SR3 always.}
\label{fig:qaoa-scaling_rzz-is-cp}
\end{figure}

Finally, we consider the results of Fig.~\ref{fig:qaoa-scaling} as an indication that better schedules are obtained by breaking the quantum algorithms in separate parts and compose the complete schedule by augmentation: when the schedule for part $k$ is obtained (possibly through a stochastic optimization as in this work), it will be used as the starting point of the schedule for part $k+1$ of the circuit. This extension will be explored in future works.

The situation completely changes when we consider a modified transmon architecture with native gates implementing ZZ rotations. In this case the scheduler can take advantage of the fact that all ZZ rotations pair-wise commute and this leads to a dramatic reduction of the number of extra SWAPs. In fact, if the circuit depth is expected to be reduced by avoiding the gate decomposition of ZZ rotations, the number of SWAPs may have been unchanged since no SWAP is required in the decomposition. However, Fig.~\ref{fig:qaoa-scaling_rzz-is-cp} shows a reduction by up to 56\% in the routing overhead from modifying the two-qubit gate order. The reduction for each number of qubits and number of QAOA layers is provided in Appendix~\ref{app:reduction-qaoa}.


\subsection{Quantum Fourier Transform}
\label{subsec:qft}

The Quantum Fourier Transform is an important subroutine used in many quantum algorithms, perhaps most notably in Shor's algorithm for prime factorization \cite{Shor1999}, in the algorithm for finding eigenvalues and eigenvectors of matrices \cite{Abrams1999}, and in quantum simulations whenever there is the necessity of moving between the position and momentum basis \cite{Zalka1998}.

The QFT gives rise to a strongly structured circuit in which each pair of qubits interact at least once according to a specific order. A very interesting point is that it has been shown that a one-dimensional array of qubits suffice for a quasi-optimal implementation of the QFT \cite{Fowler2004,Holmes2018a}, within a factor of two from the all-to-all connectivity case (under the assumption that SWAP and control-phase gates have similar cost).

\begin{figure}[b]
\centering
\includegraphics[scale=0.59]{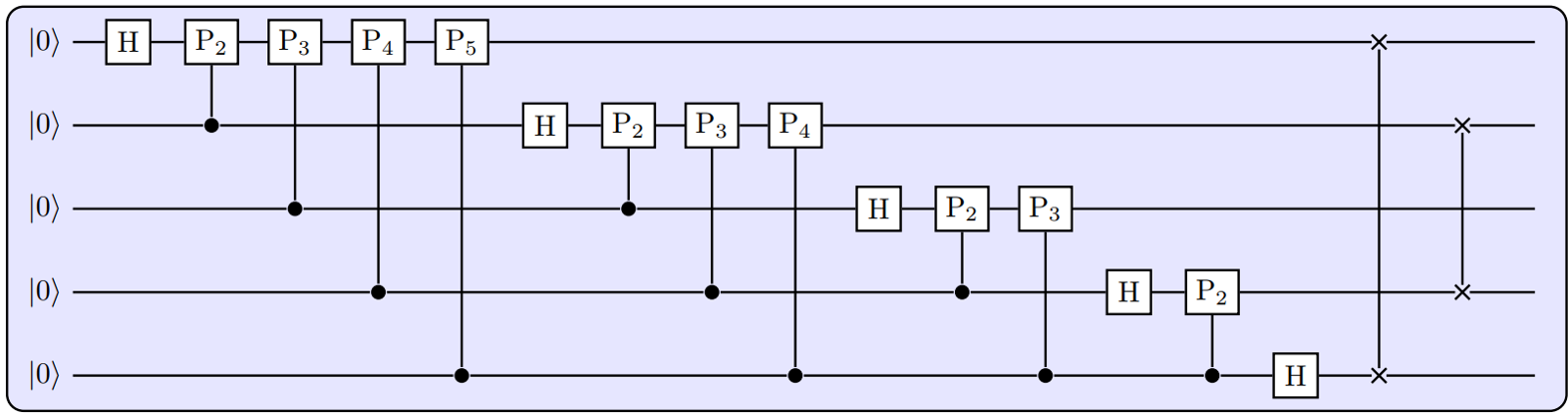}
\\ \vspace{5mm}
\includegraphics[scale=0.43]{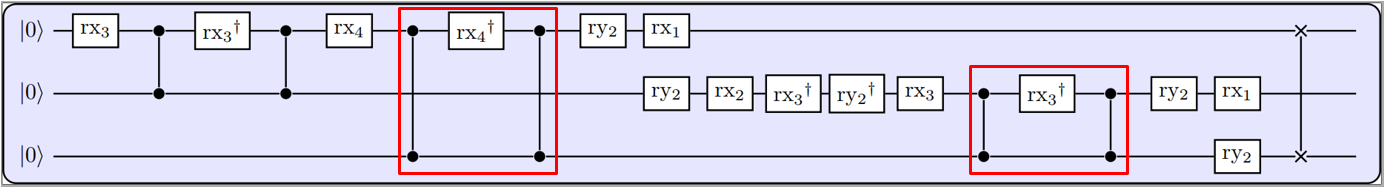}
\caption{
\textbf{Top panel:} traditional QFT circuit for 5 qubits.
\textbf{Bottom panel:} QFT circuit for 3 qubits decomposed using the native set of gates of the transmon architecture from \cite{Versluis2017}. The equivalence is modulo a non-observable global phase and the final SWAP gates can be avoided by simply updating the qubit placement. Notice that the two areas inside the red boxes   have the same priority despite being visually displaced.
\textbf{Notation:} $P_k$ is the one-qubit gate that applies the phase $\exp(i\,2\pi/2^k)$ to state $\ket{1}$ and $H$ is the Hadamard gate. For the bottom circuit, $\text{rx}_k$ is the X rotation $\text{rx}_k=\exp(-i \, 2\pi/2^k \, \sigma_x)$ generated by the Pauli matrix $\sigma_x$, and similarly for $\text{ry}_k$ generated by the Pauli matrix $\sigma_y$.}
\label{fig:qft-circuit}
\end{figure}

The general form of the QFT circuit for $n$ qubits has been presented in many works and textbooks \cite{Nielsen2011}, but we need to adapt it to the natural gate set of the architecture discussed in the previous section. In particular, we have to decompose controlled-phase gates (with different phases) into X and Y rotations and controlled-Z gates (with unique phase $\pi$). We follow the decomposition in \cite{Holmes2018a} and adapt it to the current context as visualized in Figure.~\ref{fig:qft-circuit}.

While the control constraints used in this work have been taken from two specific transmon chips \cite{Versluis2017} composed by 7 and 17 qubits arranged as a subset of a regular bi-dimensional mesh, we can generalize those constraints to other connectivities simply by indicating how the qubit frequencies are assigned and how the qubits are connected to the drive lines.

In this subsection we consider a linear array of $n$ qubits in 3 variations:
\begin{itemize}
\item[] \textbf{2 distinct frequencies}, with pattern:
$f_1 - f_2 - f_1 - f_2 - \dots - f_1 - f_2$
\item[] \textbf{3 distinct frequencies}, with pattern:
$f_1 - f_2 - f_3 - f_2 - f_1 - f_2 - f_3 - \dots - f_2 - f_1$
\item[] \textbf{all frequencies are distinct}, with pattern:
$f_1 - f_2 - f_3 - f_4 - \dots - f_{n-1} - f_{n}$
\end{itemize}
In all variations, we have a drive line per frequency group with spatial parallelism enabled. All qubits with the same frequency are connected to the same line and this allows to perform the same single-qubit operation on any subset of those qubits at the same time. However this excludes the parallel implementation of different single-qubit gates on qubits having the same frequency.

\begin{figure}[t]
\centering
\includegraphics[scale=0.51]{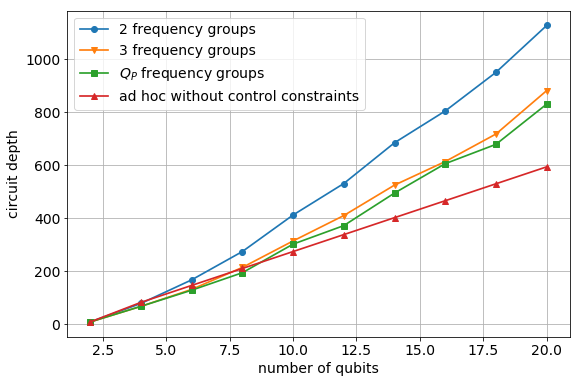}
\caption{Circuit depth of the QFT as a function of the qubit number, for transmon chips and linear connectivity. The depth is provided in clock-cycles (of 20ns for the realistic implementation of \cite{Versluis2017}) and we recall that one-qubit gates take 1 cycle, two-qubit gates take 2 cycles, SWAP gates are decomposed in a sequence of gates taking a total of 10 cycles. The purple line corresponds to analytic results in absence of control constraints, while the other datapoints have been obtained numerically using the following policies: SR0 trivial, SR1 lowest-index-first, SR2 dynamical-pattern-improvement, and SR3 always-despite-priority. Results shows the best of 10000 stochastic schedules.}
\label{fig:qft-results}
\end{figure}

We use our scheduler to analyze the effect of increasing the number of frequency groups and therefore study the impact of architectural choices on the QFT execution. The results obtained with the policy dynamical-pattern-improvement are shown in Figure.~\ref{fig:qft-results}. For comparison, we also include the results from ad-hoc compilation of the QFT circuit adapted from \cite{Holmes2018a} to the transmon gate set and gate duration. However, the ad-hoc compilation does not consider any limitation from control electronics.

Two main observations, with the first one related to the scaling behavior of the circuit depth as a function of the number of qubits $Q_L$. The ad-hoc compilation in absence of electronic constraints has linear scaling and the behavior is preserved in the optimized schedules when the control constraints are considered. This is an important observation since the number of gates in the QFT scales quadratically with $Q_L$ and therefore the degree of parallelism is also proportional to $Q_L$ to accommodate all of them in linear depth.

As expected, the depth is reduced when hardware requirements are relaxed. Here we observe a diminishing return of having more than 3 frequency groups. This effect is due to the way in which unwanted resonances are avoided during gate operation and to the fact that only the frequencies of the connected qubits matter when a two-qubit gate is scheduled. Since it is natural to expect that manufacturing chips with a larger number of frequency groups poses more challenges than having a small set of frequencies, our analysis suggests that there is limited computational advantage in having more than 3 frequencies. This consideration shines light on an important use-case for circuit schedulers beyond their natural need at runtime of quantum computations: they can be used in the design phase of novel architectures to evaluate the impact of architectural decisions like the qubit frequencies, connectivity, drive lines, and spatial/frequency parallelism.

The second observation is about the impact of the policies on the quality of the overall schedule. Initially we considered the SR3 always policy for snapshot update that limits the gates in $\N_\alpha$ to those with the maximum priority (per logical qubit). However this is a poor choice for QFT since there are several parts of the circuit with the same priority, but that one ideally wants to schedule in a specific order. In particular, consider the gates in the red boxes in the bottom panel of Fig.~\ref{fig:qft-circuit}: the left CP gates in the two boxes have the same priority and it is higher than the priority of the right CP gates. If we strictly impose to follow the decreasing priority as a criterion in fixing the gate execution, one would not perform both 2-qubit operations in a red box consecutively, but will bounce back-and-forth between the two boxes. When more qubits are considered, these kinds of conflicts increase in frequency and number (for $Q_L$ qubits, there are situations with $(Q_L-1)$ same-priority red boxes).
By simply opting for the SR3 always-despite-priority policy, the scheduler is free to break the strict priority order without violating any logical dependency.
